\documentclass[twoside]{ilcws08}
\usepackage[latin1]{inputenc}
\usepackage[dvips]{graphicx,epsfig,color}
\usepackage{wrapfig,rotating}
\usepackage{amssymb,amsmath,array}
\usepackage{url}

\begin{document}
\title{
Flavour Tag Studies with the LCFIVertex Package} %% 
\author{Roberval Walsh
\vspace{.3cm}\\
The University of Edinburgh - School of Physics and Astronomy\\
Mayfield Road 
Edinburgh EH9 3JZ United Kingdom
}

\maketitle

\begin{abstract}
In this contribution the status of the flavour tagging performance for the LDCPrime detector model and studies of the track selection parameters and effects from beam backgrounds in flavour tagging using the LCFIVertex package are presented. This work is part of an effort towards a default configuration for the ILD detector optimisation.
\end{abstract}

%%%%%%%%%%%%%%%%%%%%%%%%%%%%%%%%%%%%%%%%%%%%%%%
\section{Introduction}
In this contribution \cite{Slides} we present studies for optimisation of the flavour tag with the detector model LDCPrime\_02Sc using the LCFIVertex package \cite{Bailey:2009,Hillert:2007gt}. A sample of $e^+e^- \rightarrow Z \rightarrow q\bar q$, where $q=u,d,s,c,b$ at $\sqrt s = 91.2$ GeV generated with Pythia \cite{Sjostrand:2006za} was used. Detector simulation and event reconstruction were performed with the ilcsoft \cite{Ilcsoft} v01-03-06-p02. We also present the preliminary performance of the flavour tag with the detector model ILD\_00 to be used in the analyses of the benchmark processes for the International Linear Collider (ILC) Letters of Intent.

The LCFIVertex package provides tools for vertex reconstruction with the topological vertex finder ZVTOP \cite{Jackson:1996sy}, flavour tag using neural networks \cite{XellaHansen:2001au} and vertex charge determination \cite{Hillert:2005rp}. The LCFIvertex source codes and the trained neural networks for flavour tag are available in a CVS repository \cite{CVS} under {\it marlinreco} and {\it tagnets}, respectively.

%%%%%%%%%%%%%%%%%%%%%%%%%%%%%%%%%%%%%%%%%%%%%%%
\section{LCFIVertex parameters tuning}

\begin{wrapfigure}[19]{r}{0.5\columnwidth}
\centerline{\includegraphics[width=0.4\columnwidth]{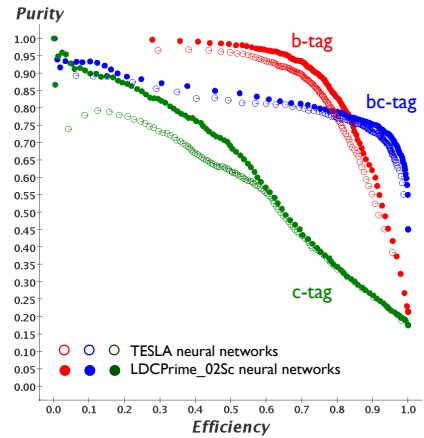}}
\caption{Flavour tagging for LDCPrime\_02Sc with neural networks obtained with TESLA and LDCPrime\_02Sc geometries.}\label{Fig:ldc_neuralnets}
\end{wrapfigure}

The flavour tag performance depends on the various parameters of LCFIVertex. We observed that it is very sensitive to the parameters for the joint probabilities \cite{Bailey:2009} that a track emerges from the primary vertex. The joint probabilities are among the input variables of the neural networks used for flavour tag. Consequently, after obtaining such parameters a re-training of the neural networks is needed. LCFIVertex has a dedicated processor (SignificanceFit) to obtain the parameters for the joint probabilities. The parameters for LDCPrime\_02Sc used in the studies are shown in Table~\ref{tab:JPpars}.

The training of the neural networks was performed using the NeuralNetTrainer processor available in LCFIVertex. The improvement in the performance of the flavour tag due to the re-training of the networks when compared with the networks trained for the TESLA design is clear from efficiency {\it versus} purity plot for the $b$-tag, $c$-tag and $bc$-tag ($c$-tag in $b$ background) shown in Figure~\ref{Fig:ldc_neuralnets}. 
\begin{wraptable}[22]{r}{0.5\columnwidth}
\centerline
{\begin{tabular}{|c|c|c|}
\hline
parameter & joint prob. & joint prob. \\
  & in $r-\phi$ & in $z$ \\
\hline
$p_1$ & $ 0.843$ & $ 0.911 $\\ 
$p_2$ & $ 0.365$ & $ 0.306$\\ 
$p_3$ & $ 0.620$ & $ 0.423$\\ 
$p_4$ & $ 0.150$ & $ 0.139$\\
$p_5$ & $ 0.029$ & $ 0.028$\\ 
\hline
 \end{tabular}}
\caption{Parameters for the joint probabilities in $r-\phi$ and $z$ for LDCPrime\_02Sc.}
\label{tab:JPpars}
\bigskip\bigskip
\centerline
{\begin{tabular}{|l|l|l|}
\hline
parameter & ipfit & zvres \\
\hline
$\chi^2/ndf$            & $< 5$          & $< 4$ \\ 
$d_0 (\rm{mm})$         & $< 20$         & $< 2$ \\ 
$d_0$ error $(\rm{mm})$ & $ - $          & $< 0.007$ \\ 
$z_0  (\rm{mm})$        & $< 20$         & $< 5$ \\
$z_0$ error $(\rm{mm})$ & $ - $          & $< 0.025$ \\ 
$p_T (\rm{GeV})$        & $> 0.1$        & $> 0.2$ \\
\hline
\end{tabular}}
\caption{Preliminary track selections for the vertices reconstruction.}
\label{tab:TrackSel}
\end{wraptable}

Flavour tag can also be sensitive to other parameters\footnote{The default values of all parameters can be found in the xml files in the directory {\it steering} of the LCFIVertex source codes.}. We looked into the track selections used to reconstruct the primary and secondary vertices. The available parameters for track selection, $r-\phi$ and $z$ impact parameters, $d_0$ and $z_0$, and their corresponding errors, the $\chi^2/ndf$ of the track fit and the transverse momentum $p_T$ of the track, were varied independently and the flavour tag evaluated. Other variables related to the vertex reconstruction were also investigated to find cuts for the above parameters keeping the vertex reconstruction performance. Preliminary optimised track selections for the primary (ipfit) and secondary (zvres) vertices reconstruction are in Table~\ref{tab:TrackSel}. Small improvements in the flavour tag performance were observed.

%%%%%%%%%%%%%%%%%%%%%%%%%%%%%%%%%%%%%%%%%%%%%%%
\section{Effects of beam backgrounds}
A strong motivation for optimising the track selection is the beam backgrounds that can degrade the performance of the track and vertex reconstruction and, consequently, of the flavour tag. To study the effects of beam backgrounds in the flavour tag, the whole event reconstruction was performed with noise hits in the vertex detector (VXD). The hits were simulated using the processor VTXNoiseHits of the MarlinReco package \cite{Ilcsoft}. The number of noise hits per cm$^2$ per layer per bunch crossing used in the simulation is $100$, $10$, $4$, $1$, $1$\footnote{During normal operation of the ILC, the predicted densities of noise hits are $400$, $50$, $15$, $6$, $3$, from the innermost to the outermost layer. For technical reasons the lower densities were used instead.} for each layer of the VXD, starting from the innermost layer. The effects on the flavour tag performance can be seen in Figure~\ref{Fig:ldc_noise}. No significant improvement was found using the track selections in Table~\ref{tab:TrackSel}.

%%%%%%%%%%%%%%%%%%%%%%%%%%%%%%%%%%%%%%%%%%%%%%%
\section{Flavour Tag with the ILD detector}
\begin{figure}[ht]
\begin{minipage}[b]{0.5\linewidth}
\centering
\includegraphics[width=0.8\columnwidth]{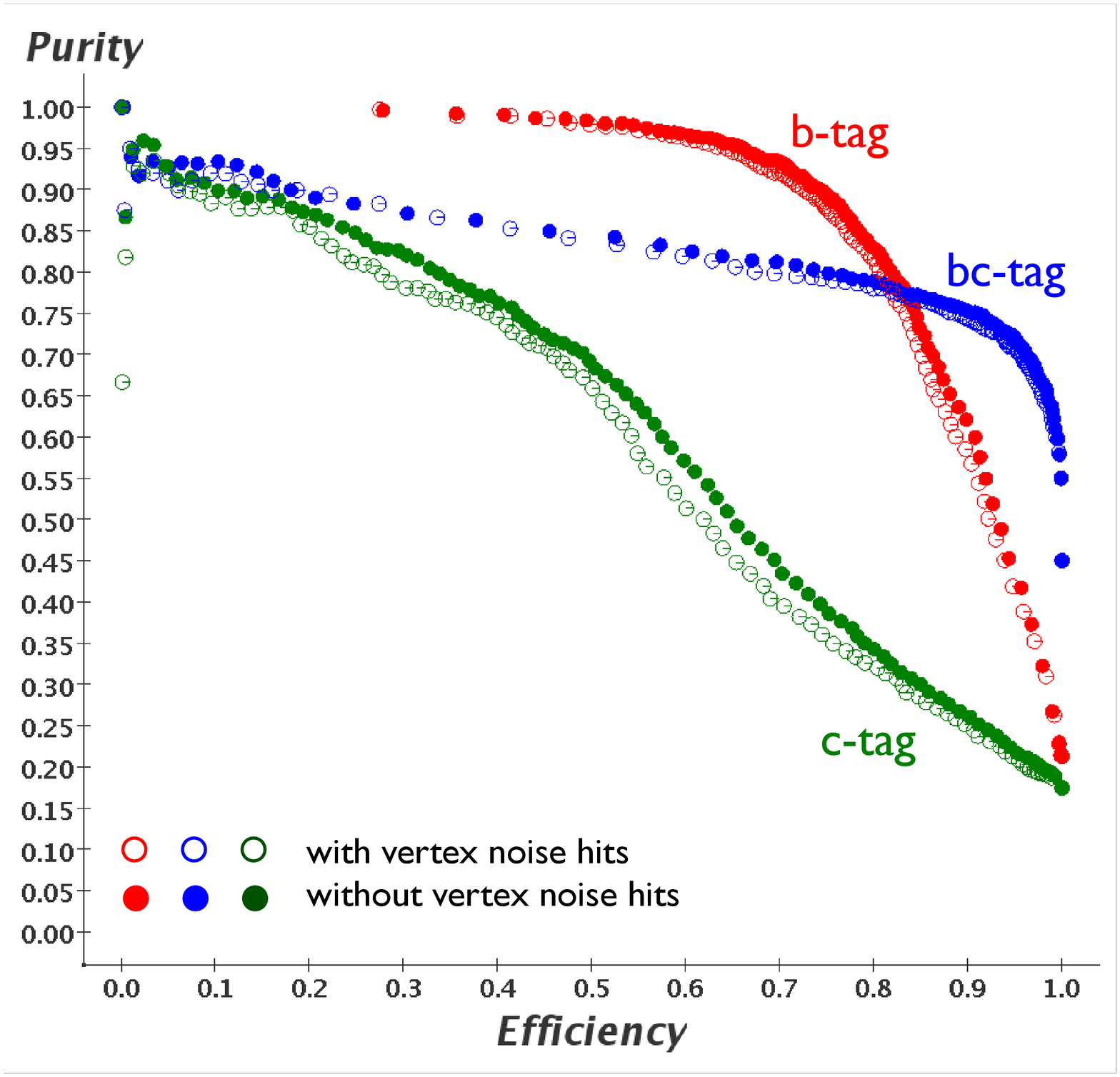}
\caption{Flavour tagging for LDCPrime\_02Sc with and without noise hits in the VXD.}
\label{Fig:ldc_noise}
\end{minipage}
\hspace{0.5cm}
\begin{minipage}[b]{0.5\linewidth}
\centering
\includegraphics[width=0.8\columnwidth]{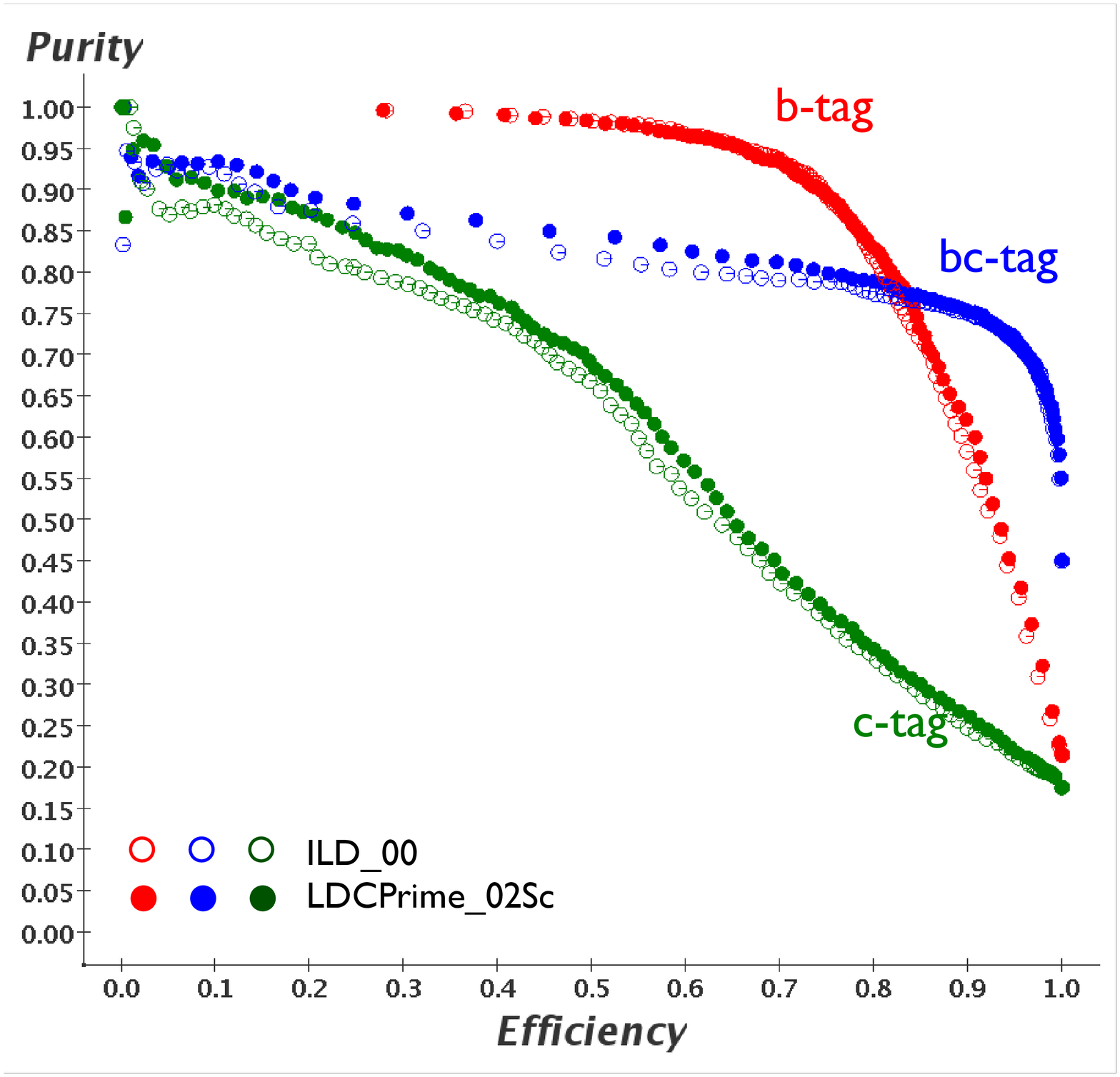}
\caption{Flavour tagging for LDCPrime\_02Sc with and without noise hits in the VXD.}
\label{Fig:ldc_ild}
\end{minipage}
\end{figure}

The ILD detector concept was born from the union of the LDC and GLD detector concepts. The choice of the geometry of the ILD VXD follows that of the GLD concept with three double layers opposed to the five single layers of the LDC concept. The performance of the flavour tag with the detector model ILD\_00 compared with LDCPrime\_02Sc can be seen in Figure~\ref{Fig:ldc_ild}. Note that no dedicated training of the neural networks was performed for ILD\_00 yet, the LDCPrime\_02Sc neural networks were used instead.

%%%%%%%%%%%%%%%%%%%%%%%%%%%%%%%%%%%%%%%%%%%%%%%
\section{Summary}
Studies of the dependence of the flavour tag on LCFIVertex parameters and the effects of the ILC beam backgrounds using the LDCPrime\_02Sc detector model were presented. For that detector model the parameters for the joint probabilities and the neural networks for the flavour tag were obtained. Track selections used in the reconstruction of the primary and secondary vertices that do not alter the performance of the vertex finding are proposed. The performance of the flavour tag does not change significantly with such selections, in particular in the presence of beam backgrounds.
Other LCFIVertex parameters are under investigation. The results presented in this contributions indicate that alternative parameters for track selections are necessary to suppress effects from beam backgrounds in the flavour tag.

\section{Acknowledgments}
This work has made use of the resources provided by the Edinburgh Compute and Data Facility (ECDF). The ECDF is partially supported by the eDIKT initiative.

% ****************************************************************************
% BIBLIOGRAPHY AREA
% ****************************************************************************
\begin{footnotesize}

\bibliographystyle{unsrt}
\bibliography{flavourtag}

\end{footnotesize}
% ****************************************************************************
% END OF BIBLIOGRAPHY AREA
% ****************************************************************************

\end{document}